\documentclass{article}
\usepackage{spconf,amsmath,graphicx}
\usepackage{url}

\usepackage{algorithm,algpseudocode,bm,amsfonts,booktabs,subfigure}
\usepackage{cite}
\usepackage{bbding}
\usepackage{multirow}
\usepackage{array}
\usepackage{bibspacing}

\title{VILAS: EXPLORING THE EFFECTS OF VISION AND LANGUAGE CONTEXT \\ IN AUTOMATIC SPEECH RECOGNITION}
\name{Ziyi Ni$^{1,2,*}$, Minglun Han$^{1,*,\dagger}$, Feilong Chen$^{1,*}$, Linghui Meng$^{1,2}$, Jing Shi$^{1,2}$, Pin Lv$^{1}$ and Bo Xu$^{1,2}$ \thanks{$*$~Co-first Authors} 
\thanks{$\dagger$~Corresponding author. This work was completed in CASIA.}}
\address{$^{1}$Institute of Automation, Chinese Academy of Sciences, Beijing, China \\
$^{2}$School of Artificial Intelligence, University of Chinese Academy of Sciences, Beijing, China \\
\small \tt \{niziyi2021, hanminglun2018\}@ia.ac.cn}

\begin{document}
\ninept
\normalsize
\maketitle
\begin{abstract}
Enhancing automatic speech recognition (ASR) performance by leveraging additional multimodal information has shown promising results in previous studies. However, most of these works have primarily focused on utilizing visual cues derived from human lip motions. In fact, context-dependent visual and linguistic cues can also benefit in many scenarios. In this paper, we first propose ViLaS (\textit{Vi}sion and \textit{La}nguage into Automatic \textit{S}peech Recognition), a novel multimodal ASR model based on the continuous integrate-and-fire (CIF) mechanism, which can integrate visual and textual context simultaneously or separately, to facilitate speech recognition. Next, we introduce an effective training strategy that improves performance in modal-incomplete test scenarios. Then, to explore the effects of integrating vision and language, we create VSDial, a multimodal ASR dataset with multimodal context cues in both Chinese and English versions. Finally, empirical results are reported on the public Flickr8K and self-constructed VSDial datasets. We explore various cross-modal fusion schemes, analyze fine-grained cross-modal alignment on VSDial, and provide insights into the effects of integrating multimodal information on speech recognition.
\end{abstract}

\section{Introduction}
\label{sec:intro}

In the traditional automatic speech recognition (ASR) task, the input is usually only a speech utterance, and the output is its corresponding transcription. However, additional cues can also help improve ASR performance. For example, the lip motions of speakers enhance the noise robustness of ASR systems~\cite{afouras2018deep}, which is suitable for video conferencing. In video subtitling, video can be used as extra visual information for better speech understanding~\cite{miao2016open,gupta2017visual}. Captions in visual dialogue~\cite{das2017visual} or video summaries can also be used as linguistic cues. Some efforts~\cite{zhao2019shallow,le2021deep} inject personal phrases into the ASR systems to customize voice assistants. 

While most works integrate a single additional modality into ASR models, the ASR systems that simultaneously input vision and language~\cite{MoriyaJ18}, have not been fully explored. There are the following challenges: 1) Cross-modal interaction. A proper modeling scheme is essential for multimodal learning, while an unreasonable scheme may make additional inputs become useless noises; 2) Missing modalities~\cite{yin2017unified,ma2021smil,ma2022multimodal}. A common assumption in multimodal learning is the completeness of the multimodal data. However, in real applications, it is not always possible to guarantee the existence of each modality of the multimodal data pairs, which may cause severe degradation;
3) Dataset construction. Compared with text data, visual data and audio data are more difficult to collect. Therefore, it is challenging to create high-quality paired data from scratch. Besides, a good multimodal dataset is supposed to fully reflect the potential impact of any additional integrated modalities. 

Aiming at cross-modal modeling, we first propose a multimodal ASR model (ViLaS), which can extract high-level acoustic features with the speech encoder based on the continuous integrate-and-fire mechanism (CIF)~\cite{dong2020cif} and can incorporate multiple additional modalities simultaneously or a single additional modality. 
We then introduce an effective training strategy to improve the basic acoustic modeling capability of ViLaS through \textbf{pre-training}, and its ability to handle missing-modal test scenarios through \textbf{mixed-training}. 
Next, to better explore the multimodal effects for ASR tasks, we construct a multimodal ASR dataset called visual spoken dialogue (VSDial) based on VisDial~\cite{das2017visual}.
VSDial covers speech, visual context and textual context, and has two versions in English and Chinese. 
Finally, we experiment with our methods on Flickr8K and VSDial, and find that both visual and linguistic cues can bring improvements for the ASR on VSDial. 
It is worth noting that the CIF mechanism allows us to visualize the cross-modal alignment between speech frames, tokens and images, which provides insights into the inference behaviors.

\section{Related work}
\label{sec:relate work}

\textbf{Integrating visual information into ASR.}
Visual cues in previous research are mainly derived from lip movement~\cite{xu2020discriminative,Shi2022RobustSA}, requiring strict synchronization with audio frames and precise real-time visual localization~\cite{gupta2017visual}.
Solely introducing lip position may result in the loss of background details for context understanding. Meanwhile, the exploration of context-dependent visual information, also known as visual context, is limited. In recent years, several studies~\cite{Sun2016LookLA, Srinivasan2020MultimodalSR,Srinivasan2020FineGrainedGF,GhorbaniGSL21} have incorporated visual contexts, such as images or videos, into ASR models with attention mechanisms and shown its effectiveness. ~\cite{Pramanick2022CanVC} extracts visual context by a visual object detector for embodied agents. Some studies~\cite{GhorbaniGSL21} found that visual context improves the recognition of nouns, while others~\cite{Srinivasan2020FineGrainedGF} analyzed the effect of vision on different linguistic components.

\noindent\textbf{Integrating linguistic information into ASR.} 
Various techniques are proposed for integrating external language models into end-to-end ASR models, such as shallow fusion\cite{toshniwal2018comparison,kannan2018analysis}, deep fusion~\cite{toshniwal2018comparison}, and cold fusion\cite{sriram2017cold}. Some recently leverage the power of large pre-trained language models through knowledge distillation~\cite{futami2020distilling,kubo2022knowledge,han2023knowledge} and re-scoring~\cite{Xu2022RescoreBERTDS}. Another line of work aims to incorporate knowledge in the form of context-specific phrases or dialogue context within the original ASR framework. 
For example, CLAS~\cite{pundak2018deep}, Contextual RNN-T~\cite{Jain2020ContextualRF}, and ColDec~\cite{han2021cif,han2022improving} integrate contextual phrases to achieve customization. Similarly, ~\cite{DM19,Hou2022BringDI} apply attention-based fusion to integrate dialogue context.

\noindent\textbf{Integrating multimodal information into ASR.}~\cite{MoriyaJ18} adopted a pipeline for integrating images and titles, which comprises an acoustic model and LSTM language model.
\section{Proposed methods}
\label{sec:methods}
\subsection{ViLaS}

\begin{figure}[t]
  \centering
    \vspace{-5pt}
    \includegraphics[width=1\linewidth]{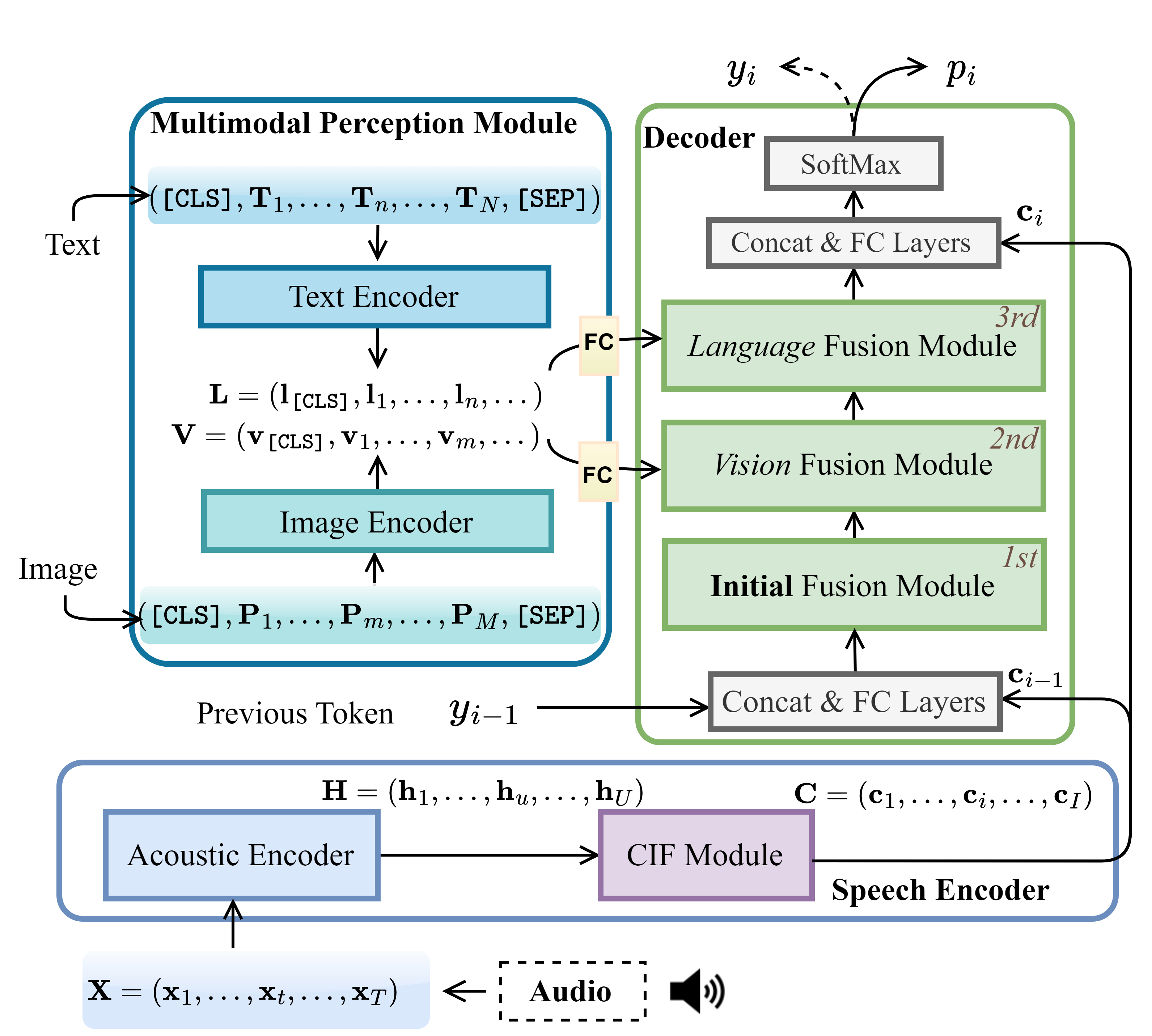}
  \caption{Our general multimodal ASR framework: ViLaS.}
  \label{fig:vilas}
  \vspace{-15pt}
\end{figure}

Here, we design a generic multimodal ASR framework, 
which can integrate multiple modalities simultaneously or only a single modality with the help of cross-attention,
to capture context information related to speech for better understanding. 
We present ViLaS, a multimodal ASR model built on the continuous integrate-and-fire (CIF) mechanism~\cite{dong2020cif}. As shown in Figure~\ref{fig:vilas}, it has three primary modules: \textbf{the speech encoder} includes an acoustic encoder and a CIF module, and encodes raw speech features into the high-level acoustic sequence; \textbf{the multimodal perception module} encodes vision and language with text and image encoders; \textbf{the decoder} consists of three fusion modules and several fully-connected layers (FC Layers), performs decoding by consuming features in high-level acoustic sequence one-by-one, and incorporates multimodal features at each decoding step via the attention mechanism.

\textbf{Speech Encoder.} The raw speech features is $\mathbf{X}$ extracted from the audio. It is first fed to the acoustic encoder, and then its output $\mathbf{H}$ is delivered to the CIF module. Next, the CIF module outputs high-level acoustic sequence $\mathbf{C}$. Specifically, the CIF firstly accumulates the weight of each frame from $\mathbf{H}$, then determines the acoustic boundaries of adjacent tokens by judging whether the accumulated weight exceeds the pre-defined threshold, and finally summarizes the features within two adjacent acoustic boundaries in $\mathbf{H}$ via weighted sum to obtain $\mathbf{C}$. Thus, the high-level acoustic sequence $\mathbf{C}$, which is strictly aligned with the transcription, can also be regarded as a non-uniformly compressed version of the low-level acoustic sequence $\mathbf{H}$ via variable-length sub-sampling.

\textbf{Multimodal Perception Module.} The multimodal perception module contains the image encoder and text encoder. We specially introduce a pre-trained vision transformer (ViT)~\cite{DosovitskiyB0WZ21} as the image encoder. We split the original image into fixed-size patches. ViT takes the patch representation sequence as the input, and output a feature sequence $\mathbf{V}$. The text encoder is a pre-trained language model called BERT~\cite{DevlinCLT19}. Given the input text sequence, the output of BERT is a feature sequence $\mathbf{L}$. 
An extra FC layer is used to project input features to match the hidden size of the decoder for each modality. 
Both encoders are fixed during training.

\textbf{Decoder.} The decoder utilizes the high-level acoustic feature sequence $\mathbf{C}$, along with $\mathbf{V}$ and $\mathbf{L}$. Specifically, at the $i$-th decoding step, the decoder first concatenates the high-level acoustic feature $c_{i-1}$ with the embedding of the previously predicted token $y_{i-1}$, and then feeds the combined feature to the initial fusion module through an FC layer. The initial fusion module contains two transformer blocks. The visual and language fusion modules both consist of 2 transformer blocks with cross-attention, taking the output of the previous module as the query and multimodal cues (the visual or linguistic features: $\mathbf{V}$ or $\mathbf{L}$) 
as keys/values for the attention. 
Finally, 
the output of the last fusion module is combined with the high-level acoustic feature $c_{i}$ of the current step, to predict the next output token $y_{i}$. 
$p_{i}$ presents the probability distribution of the token vocabulary. 

It is particularly noteworthy that 1) in actual implementations, the integrated positions of visual and language fusion modules are adjustable, and 2) when not using additional visual and linguistic cues, we set $\mathbf{V}$ to $(\mathbf{0})$, and set $\mathbf{L}$ to $(\mathbf{0})$. 
\subsection{Training strategy}

Our main objective is to propose training strategies aimed at enhancing multimodal speech recognition (MSR) performance using general-purpose ASR data. Collecting MSR high-quality data can be challenging. 
Therefore, 
it is crucial to fully utilize relatively higher-quality and larger-scale generic data to augment the foundational modeling capabilities in MSR. 
Besides, if the model is trained solely on delicately constructed paired data, even with the masking strategies inspired by the visual research to balance attention across modalities~\cite{gabeur2022avatar}, it is still heavily reliant on the specific modal inputs presented during training, resulting in overfitting in practice. 
That is, the model trained in full-modality settings may struggle to generalize and make accurate predictions when faced with missing modalities during inference.

A training strategy is introduced to improve the basic modeling ability of ViLaS by pre-training, and to handle the problem of missing modalities by mixing up generic data and multimodal data. Our training strategy contains two training phases: \textbf{Pre-training phase}: Train a basic generic ASR model $M_a$ with the generic ASR dataset $D_a$. \textbf{Mixed-training phase}: Mix up a generic ASR dataset $D_1$ (can come from from $D_a$) and a multimodal ASR dataset $D_2$, and denote the mixed dataset as $D_m$. Initialize the speech encoder of a multimodal ASR model $M_m$ with the parameters of the pre-trained basic generic model $M_a$. Train the $M_m$ with the mixed dataset $D_m$. We describe details about creating $D_m$ in section~\ref{detail_data}.

\section{Experimental setup and Results}

\subsection{Datasets}
\label{detail_data}
We use two \textbf{generic ASR datasets}: AISHELL-1 for Chinese~\cite{bu2017aishell}, and LibriSpeech for English~\cite{Panayotov2015LibrispeechAA}, and two \textbf{multimodal ASR datasets}: a public dataset Flickr8K and a self-constructed dataset called visual spoken dialogue (VSDial). 

During the pre-training phase, $M_a$ is pre-trained on the full training set of the generic ASR dataset $D_a$. During the mixed-training phase, we first denote the number of training utterances of the chosen generic dataset and multimodal dataset as $N_a$ and $N_m$. Then, we sample the same $K=min(N_a, N_m)$ utterances from the training data of the generic and multimodal datasets denoted as $D_1$ and $D_2$, respectively. Finally, $D_1$ and $D_2$ are mixed as $D_m$ containing $2K$ utterances.

\textbf{Flickr8K.} 
Flickr8K~\cite{Hodosh2013FramingID} consists of 8k images, each paired with five different captions recorded in English~\cite{Harwath2015DeepMS}. Thus, Flickr8K comprises 30k audio training samples, with 5k each in the validation and test sets. We use the image (as the visual context) to help recognize the speech utterance of the caption.

\textbf{VSDial.} We create the visual spoken dialogue (VSDial) dataset from the visual dialogue (VisDial)~\cite{das2017visual}, which includes approximately 120k images, each accompanied by 1 caption and 10 rounds of dialogue comprising a question and an answer. In multimodal ASR, we record the questions in dialogues for speech, and use the image as the visual context, and the caption as the textual context to help recognize the question speech. Thus, VSDial includes roughly 1.2 million multimodal ASR training samples. 

We create two versions of VSDial: VSDial-CN and VSDial-EN for Chinese and English, respectively. For \textbf{VSDial-EN}, all questions in the VisDial are first synthesized into speech with FastSpeech2~\cite{0006H0QZZL21} trained on LJSpeech~\cite{ljspeech17}. Regarding the creation of our test sets, we extract 2,000 question-answer pairs from the VisDial test set, which we refer to as ``test-orig" in our work. Additionally, following the questioning style employed in VisDial, we artificially generate 300 questions based on images from the VisDial test set, creating the VSDial test set called ``test-art". The speech utterances of test sets are recorded with Android phones or iPhones. For \textbf{VSDial-CN}, we first translate all the training text in VisDial into Chinese with the machine translation model\footnote{{https://huggingface.co/facebook/mbart-large-50-many-to-many-mmt}}, and then synthesize all questions into speech using FastSpeech2 trained on AISHELL-3~\cite{shi2020aishell}. To create our Chinese test sets, we translate the ``test-orig" and ``test-art" of the VSDial-EN into Chinese, and manually fix translation errors, record the speech utterances with Android phones or iPhones. All recorded audios are finally re-sampled to 16kHz. Note that we use the synthesized version of the VisDial dev set as the VSDial dev set. More details about test sets in VSDial are in Table~\ref{dataset_testset}.

\vspace{-3pt}
\subsection{Configurations}

We extract 80-channel filter-banks features computed from a 25ms window with a stride of 10ms. For Chinese, the output vocabulary contains 4,230 characters and four extra tokens \texttt{[PAD]}, \texttt{[EOS]}, \texttt{[BOS]}, \texttt{[UNK]}. For English, the output vocabulary contains 4,996 tokens generated using BPE~\cite{sennrich-etal-2016-neural} and those four extra tokens. As for metrics, we use character error rate (CER) for Chinese and word error rate (WER) for English, which is widely used in the context of ASR. 
For Chinese experiments, the speech encoder of ViLaS consists of a convolution front-end and a conformer module~\cite{gulati2020conformer}. The convolution front-end is a 2-dimensional convolution layer with 128 output channels, kernel size 3, and stride 2. The conformer module consists of 15 conformer blocks with $d_{model}=256$, $d_{ffn}=2048$, $n_{head}=4$ 
and kernel size 15 (for depth-wise convolution), and 2 max-pooling layers after the 5th and the 10th blocks. The CIF module contains a 1-dimensional convolution layer with 256 output channels, kernel size 3 and stride 1, and an FC layer followed by a $\operatorname{sigmoid}$ activation. The decoder consists of several FC layers and three transformer fusion modules comprised of 6 transformer blocks with $d_{model}=256$, $d_{ffn}=2048$ and $n_{head}=4$. In the original order, the 1st and the 2nd blocks, the 3rd and the 4th blocks, and the 5th and the 6th blocks are for the initial fusion (1st module), vision fusion (2nd module), and language fusion (3rd module), respectively. 
We will further discuss the effect of the specific order of the vision and language fusion module in Table~\ref{tab:struc}.
For English experiments,  $d_{model}$ and $n_{head}$ is changed to 512 and 8 respectively for the decoder, while others are the same as the Chinese. ViT and BERT are from transformers~\cite{transformers}\footnote{https://huggingface.co/openai/clip-vit-base-patch16, https://huggingface.co/bert-base-uncased, https://huggingface.co/bert-base-chinese}.

During training, we apply dropout for the conformer and transformer. We use SpecAugment~\cite{park2019specaugment} and label smoothing with $\epsilon=0.1$. We train models with the Adam optimizer~\cite{kingma2014adam} with $lr=\operatorname{3e-4}$ and weight decay of 1e-2. 
The training objective is the same as the original CIF-based model~\cite{dong2020cif}. The weights of CE loss, CTC loss, and quantity loss are 1.0, 0.5, and 1.0. In reference, we use beam size 10 for beam search.

\begin{table}[t]
  \centering
  \caption{Details of the test sets in VSDial dataset.}
    \vspace{5pt}
    \renewcommand\arraystretch{0.9}
    \scalebox{1}{
    \resizebox{\linewidth}{!}{
    \begin{tabular}{l|ccccc}
    \toprule
    {Dataset} & {Name} & {\# Speaker} & {\# Utterance} & {Gender} & {Source} \\
    \midrule
    \multirow{2}[2]{*}{{VSDial-CN}} & {test-orig}  & 1   & 2000  & male & VisDial testset \\
          & {test-art} & 6 & 300 & all & Manually Generated\\
    \midrule
    \multirow{2}[2]{*}{{VSDial-EN}} & {test-orig}  & 2    & 2000  & all & VisDial testset\\
          & {test-art} & 2 & 300 & all & Manually Generated\\
    \bottomrule
    \end{tabular}%
    }
    }
  \label{dataset_testset}%
  \vspace{-14pt}
\end{table}%

\begin{table}[t]
  \centering
  \caption{Main WER($\%$) results on test of F8K and test-clean of LS.}
  \vspace{5pt}
      \renewcommand\arraystretch{0.7}
    \scalebox{0.8}{
    \resizebox{\linewidth}{!}{
    \begin{tabular}{ll|cc|cc|c}
    \toprule
\multirow{2}[4]{*}{ID} & \multirow{2}[4]{*}{{Model}} & \multicolumn{2}{c|}{{Training}} & \multicolumn{2}{c|}{{F8K}} & {LS} \\
\cmidrule{3-7}          &       & {w/ V} & {Data} & {w/o V} & {w/ V} & {w/o V} \\
    \midrule
    \texttt{A1} &~\cite{Srinivasan2020FineGrainedGF} & \XSolidBrush  & $D_2$ & 13.6 & - & \\
    \texttt{A2} &~\cite{Srinivasan2020FineGrainedGF}  & \Checkmark     & $D_2$ & - & 14.1 &  \\
    \midrule
    \texttt{B1} &~\cite{oneațua2022improving} (\textbf{PT})  &  & $D_a$ & 11.1 & - & 2.6 \\
    \texttt{B2} &~\cite{oneațua2022improving} (PT+\textbf{FT})  & \XSolidBrush   & $D_2$ & 3.8 & - &  \\
    \texttt{B3} &~\cite{oneațua2022improving} (PT+\textbf{FT}) & \Checkmark     & $D_2$ & - & 4.3 &  \\
    \midrule
    \texttt{V1} & {CIF-based (\textbf{PT})} &      & $D_a$ & 11.0 & - & 3.2 \\
    \texttt{V2} & {ViLaS}  & \XSolidBrush     & $D_2$ & 11.0 & - & 85.9 \\
    \texttt{V3} & {ViLaS} & \Checkmark       & $D_2$ & 12.3 & 11.0 & 86.5 \\
    \texttt{V4} & {ViLaS (PT+\textbf{FT})}  & \XSolidBrush  & $D_2$ & 4.4 & - & 50.6 \\
    \texttt{V5} & {ViLaS (PT+\textbf{FT})} & \Checkmark    & $D_2$ & 4.7 & 4.5 & 52.5 \\
    \texttt{V6} & {ViLaS (PT+\textbf{FT})} & \XSolidBrush  & \textbf{$D_m$} & 3.4 & - & 4.0 \\
    \texttt{V7} & {ViLaS (PT+\textbf{FT})} & \Checkmark    & \textbf{$D_m$} & 3.4 & 3.4 & 4.1 \\
    \bottomrule
    \end{tabular}%
    }
}
  \label{tab:flkr8k_main}%
    \vspace{-14pt}
\end{table}%

\begin{table}[t]
  \centering
  \caption{Under different training settings, main CER($\%$) results on VSDial-CN and WER($\%$) results on VSDial-EN. ``w/o M", ``w/ M", ``w/ V" and ``w/ L"  means without any multimodal cues, with multimodal cues, with only vision and only language, respectively. The meaning of "w/M" for the test is consistent with the training stage.}
  \vspace{5pt}
    \renewcommand\arraystretch{1}
 \scalebox{1}{
  \resizebox{\linewidth}{!}
   { \begin{tabular}
   {ccccc|cc|cc|cc|cc p{1.1cm}<{\centering}|p{1.2cm}<{\centering}|p{1.2cm}<{\centering}}
   
    \toprule
    && \multirow{2}{*}{Training} && & \multicolumn{4}{c|}{VSDial-CN} & \multicolumn{4}{c}{VSDial-EN} \\
    \cmidrule{6-13} &&&&&\multicolumn{2}{c|}{test-orig} & \multicolumn{2}{c|}{test-art}  & \multicolumn{2}{c|}{test-orig} & \multicolumn{2}{c}{test-art} \\
\cmidrule{1-13} PT & w/ M & w/ L &w/ V & Data  & w/o M & w/ M  & w/o M & w/ M & w/o M & w/ M & w/o M & w/ M\\

    \midrule
    \XSolidBrush & \Checkmark & \Checkmark & \Checkmark & $D_2$ & 26.9 & 22.4 & 29.1 & 26.1 & 73.8 & 69.9 & 77.2 & 72.8 \\
    \Checkmark & \Checkmark & \Checkmark & \Checkmark & $D_2$ & 6.3 & 5.1 & 11.7 & 10.9 & 32.4 & 31.6 & 32.8 & 32.1 \\
    \Checkmark & \XSolidBrush & \XSolidBrush & \XSolidBrush & $D_m$ & 5.5  & - & \textbf{\textit{8.7}} & - &  22.5 & - & 21.8 & - \\
    \Checkmark & \Checkmark & \Checkmark & \Checkmark & $D_m$ & 6.1 & \textbf{4.7} & 9.5 & \textbf{8.5} & 22.6 & \textbf{21.3} & 23.0 & \textbf{21.5} \\
     \Checkmark & \Checkmark & \Checkmark & \XSolidBrush & $D_m$ & 6.1  & 5.1 & 9.8 & 8.6 &  22.4 & 21.0 & 21.7 & 19.5 \\
    \Checkmark & \Checkmark & \XSolidBrush & \Checkmark  & $D_m$ & \textbf{\textit{5.4}}  & \textbf{\textit{4.4}} & 9.1 & \textbf{\textit{8.3}} & \textbf{\textit{22.4}} & \textbf{\textit{21.0}} & \textbf{\textit{21.7}} & \textbf{\textit{19.5}} \\
    \bottomrule
    \end{tabular}%
    }
    }
  \label{tab:main_vsdial}%
  \vspace{-15pt}
\end{table}%

\begin{table}[t]
  \centering
  \caption{Exploration of different fusion schemes on VSDial-CN. 
  The second and third column shows the integrated position of vision and language for fusion modules, respectively.
  }
  \vspace{5pt}
  \renewcommand\arraystretch{1.0}
  \scalebox{0.6}{
    \resizebox{\linewidth}{!}{
    \begin{tabular}{l|cc|ccc}
    \toprule
    \multirow{2}{*}{ID} & \multicolumn{2}{c|}{The integrated position} & \multicolumn{3}{c}{CER on VSDial-CN} \\
    \cmidrule{2-6}          & {vision} & {language} & {test-orig} & {test-art} & {Average} \\ 
    \midrule
    \texttt{F1} & {2nd}  & {3rd}   & 4.7   & 8.5   & 6.6 \\
    \texttt{F2} & {3rd}  & {2nd}   & 5.4   & 8.7   & 7.1 \\
    \midrule
    \texttt{F3} & {2nd}   & -     & \textbf{4.4}   & \textbf{8.3}   & \textbf{6.3} \\
    \texttt{F4} & {3rd}   & -     & 4.7   & 9.6   & 7.2 \\
    \texttt{F5} & {2nd, 3rd} & -   & 4.6   & 8.3   & 6.5 \\
    \midrule
    \texttt{F6} & -     & {2nd}   & 4.8   & 9.0   & 6.9 \\
    \texttt{F7} & -     & {3rd}   & 5.1   & 9.0   & 7.0 \\
    \texttt{F8} & -     & {2nd, 3rd} & 5.1   & 8.3   & 6.7 \\
    \bottomrule
    \end{tabular}%
    }
}
  \label{tab:struc}%
  \vspace{-10pt}
\end{table}%

\begin{figure}[t]
  \centering
  \includegraphics[scale=0.38]{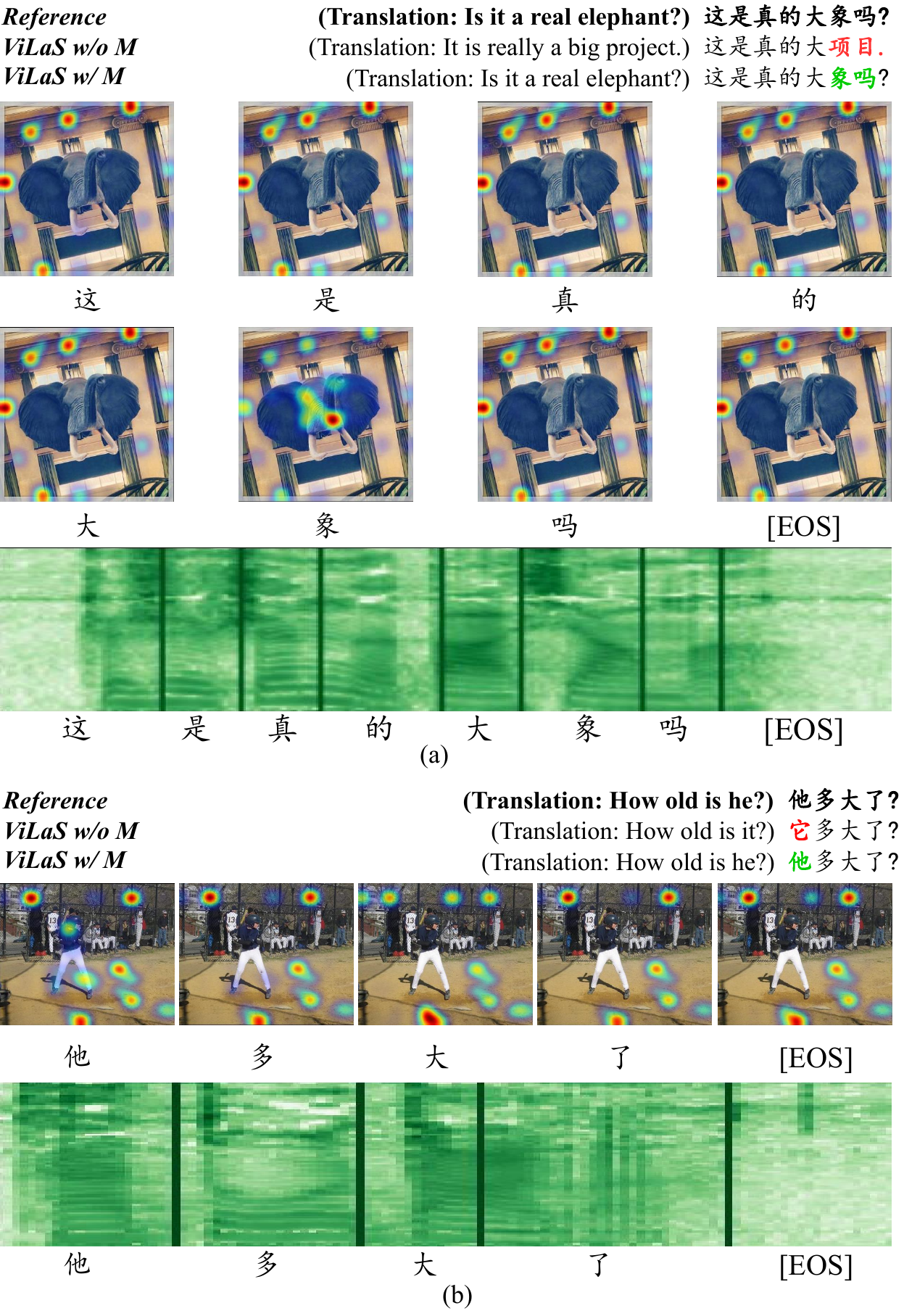} 
    \vspace{-15pt}
  \caption{The visualization of cross-modal alignment between tokens, speech frames, and the image. On the top of each figure shows the reference, the hypothesis of decoding w/o M and w/ M. The alignment between tokens and speech frames is generated according to the acoustic boundaries from the CIF.}
  \label{fig:case}
  \vspace{-10pt}
\end{figure}

\vspace{-3pt}
\subsection{Results on Flickr8K}

In Table~\ref{tab:flkr8k_main}, PT means pre-training, FT means fine-tuning. When \textbf{Model} is marked with ``(PT+FT)", the description of \textbf{Training} refers to the FT process. ``w/ V" and ``w/o V" means with and without visual inputs, respectively. F8K is Flickr8K. LS refers LibriSpeech to verify the generic ASR ability. 
From Table~\ref{tab:flkr8k_main}, both \cite{Srinivasan2020FineGrainedGF} (\texttt{A}-series) and~\cite{oneațua2022improving} (\texttt{B}-series) show that the utilization of visual cues can not yield further improvements. 
In~\cite{oneațua2022improving}, a generic ASR model pre-trained on LS ($D_a$) is used to initialize the multimodal model, which significantly boosts the performance on F8K.
\texttt{V}-series show the results of our methods. 
\texttt{V1} presents the performance of the basic $M_a$ trained with LS ($D_a$)~\cite{dong2020cif}. \texttt{V2} and \texttt{V3} present the results of training ViLaS on F8K w/ V and w/o V. In \texttt{V3}, visual cues cannot bring any improvements. 
\texttt{V4} and \texttt{V5} fine-tune ViLaS (initialized by pre-trained basic model $M_a$ in \texttt{V1}) on F8K w/ V and w/o V, while \texttt{V6} and \texttt{V7} use mixed $D_m$ w/ V and w/o V for fine-tuning.
Likewise, the comparison (\texttt{V4} vs. \texttt{V5} and \texttt{V6} vs. \texttt{V7}) shows that visual cues bring no extra gains, which is consistent with previous studies~\cite{Srinivasan2020FineGrainedGF, oneațua2022improving}. We think the reason may be that the main input modality to the ASR model is speech, and other modalities are more likely to play the role of auxiliary cues. When the speech is clear enough, other modalities may disturb recognition and cause degradation. Comparing \texttt{V3} and \texttt{V5}, we find that pre-training significantly improves ASR ability. Comparing \texttt{V5} and \texttt{V7}, we find that mixed-training makes ViLaS better adapt to the test data w/o V and alleviates the modality incompleteness problem.
\subsection{Results on VSDial}

Table~\ref{tab:main_vsdial} shows the VSDial results, which demonstrate the efficacy of pre-training and mixed-training. When fine-tuning ViLaS using $D_m$ on VSDial-CN, integrating all multimodal cues reduces CER from 5.5\% to 4.7\% on test-orig and from 8.7\% to 8.5\% on test-art. Similar improvements are also observed on VSDial-EN. From Table~\ref{tab:main_vsdial}, we can conclude that 1) when integrated alone, both vision and language could bring gains; 2) vision can bring more gains than language; 3) integrating them together could provide improvements but not cumulative gains, which implies that our cross-modal interaction scheme still has room for future explorations.

As shown in Table~\ref{tab:struc}, we conduct experiments on VSDial-CN to explore different fusion schemes. The ``2nd" or ``3rd" means the ``2nd" or ``3rd" fusion module integrated extra cues. \texttt{F3} has the best result. By comparing \texttt{F3} with \texttt{F4}, \texttt{F6} with \texttt{F7}, we find that at the earlier stage the extra cues are integrated, the more significant the improvement is.
The comparison between \texttt{F3} and \texttt{F6} verifies the superiority of vision over language.

As shown in Figure~\ref{fig:case}, with the help of the CIF mechanism, ViLaS provides alignment between tokens, frames, and the image for cases of VSDial-CN. At each step (for each token), the decoder captures context information from the image and generates corresponding attention maps. In case (a), the token of the 6th step means ``elephant" in English. Its attention map exactly assigns much attention to the ``elephant" region, while other steps whose tokens are adverbs, adjectives, and interjections mainly attend to the background. 
Compared to ViLaS w/o M, ViLaS w/ M can avoid errors by referring to visual context, which confirms the positive effect of visual inputs. In case (b), since the third-person pronouns (he/she/it) in Chinese share the same pronunciation, the model cannot easily discriminate them without additional context, thus misidentifying ``he" as ``it". Because the vision injected provides gender information, the model can capture the cross-modal alignment and correct the substitute error. The above cases demonstrate that multimodal cues can bring gains for recognizing context-dependent nouns or pronouns. 

\section{Conclusions}
We propose a novel multimodal ASR model, a set of training strategies, and a dataset (VSDial) that incorporates additional visual and linguistic cues beyond speech. We conduct experiments on both the Flickr8K and VSDial datasets under various conditions, exploring different cross-modal fusion schemes, and analyzing cross-modal alignment. Moving forward, we plan to continue developing cross-modal fusion techniques for more effective cross-modal interaction.

\section{Acknowledgments}
The authors would like to thank Shuang Xu for her previous assistance. This work was supported by the National Key R\&D Program of China under Grant No. 2020AAA0108600.

\vfill\pagebreak
\bibliographystyle{IEEEbib}
\bibliography{strings,refs}


\end{document}